\documentclass[nonblindrev]{informs3}
\usepackage{balance}
\usepackage{amsmath}
\usepackage[tight,footnotesize]{subfigure}
\usepackage{graphicx}
\usepackage{bm}
\usepackage{algorithm}
\usepackage[noend]{algorithmic}
\usepackage{multirow}
\usepackage{slashbox}
\usepackage{caption}
\usepackage{amsfonts}

\OneAndAHalfSpacedXII 

\usepackage{natbib}
 \bibpunct[, ]{(}{)}{,}{a}{}{,}%
 \def\bibfont{\small}%
\TheoremsNumberedThrough     

\EquationsNumberedThrough    

\MANUSCRIPTNO{2015} 

\begin{document}


\RUNAUTHOR{Tang et al.}

\RUNTITLE{Going viral: Optimizing Discount Allocation in Social Networks for Influence Maximization}

\TITLE{Going viral: Optimizing Discount Allocation in Social Networks for Influence Maximization}

\ARTICLEAUTHORS{%
\AUTHOR{Shaojie Tang}
\AFF{Naveen Jindal School of Management, The University of Texas at Dallas}
\AUTHOR{Jing Yuan}
\AFF{Department of Computer Science, The University of Texas at Dallas}
} 

\ABSTRACT{In this paper, we investigate the discount allocation problem in social networks. It has been reported that 40\% of consumers will share an email offer with their friend and 28\% of consumers will share deals via social media platforms. What does this mean for a business? Essentially discounts should not just be treated as short term solutions to attract individual customer, instead, allocating discounts to a small fraction of users (called seed users) may  trigger a large cascade in a social network. This motivates us to study the influence maximization discount allocation problem: given a social network and budget, we need to decide to which initial set users should offer the discounts, and how much should the discounts be worth. Our goal is to maximize the number of customers who finally adopt
the target product. We investigate this problem under both non-adaptive and adaptive settings. In the first setting, we have to commit the set of seed users and corresponding discounts all at once in advance. In the latter case, the decision process is performed in a sequential manner, and each seed user that is picked provides the feedback on the discount, or, in other words, reveals whether or not she will adopt the discount. We propose a simple greedy policy with an approximation ratio of $\frac{1}{2}(1 - 1/e)$ in non-adaptive setting. For the significantly more complex adaptive setting, we propose an adaptive  greedy policy with bounded approximation ratio in terms of expected utility.}


\maketitle

%

\section{Introduction}\label{sec:introduction}
With the rapid expansion of World Wide Web in the last two decades, social networks are becoming important dissemination and marketing platforms as they allow the efficient generation, dissemination, and sharing of information and ideas. And platforms utilizing social media have been recognized as revolutionizing
communication channels for corporations and consumers alike.  Consider the following scenario. Suppose a firm would like to generate demand for a
new product through a social network, they may choose to provide discounts to a few selected users in the network, and hope that they promote this product to their friends as well. This is different from traditional discount allocation strategy whose central problem
is to find the ``best fit'' between a discount and a given user while ignoring the network effect of that user. Our work fundamentally differs
from existing works as we are concerned with the \emph{network value} of users when making
allocations. Suppose the total discount is constrained by a budget defined by the firm, this raises the question: given the structure of the social networks and the knowledge of  how the new
product adoption diffuse through the network, which initial set users should be selected to receive the discount? and how much should the discounts be worth?

To this end, we formulate and investigate the discount allocation problem as follows. Given a social network  $G = (V, E)$, where $V$ is a set of individuals and $E$ is a set of social ties. Our model decomposes the cascade process into two stages: \emph{seeding stage} and \emph{diffusion stage}. In the seeding stage, we offer discounts to a set of initial users. Each type of discount stands for a specific dollar amount off of the purchase. Some initial users accept the offer and act as starting points, called \emph{seeds},  in the diffusion stage. It is reasonable to assume
that the adoption probability of any initial user is monotonically increasing with respect to discount rate. In the diffusion stage, the adoption propagates from the seeds to other users.  In order to model the diffusion dynamics in a social network, we can leverage the existing results done in diffusion of information  in social networks. In particular, we adopt \emph{independent cascade model} \citep{kempe2003maximizing}, which is one of the most commonly used models.  Our goal is to find the optimal configuration, which consists of a set of initial users and discount rate for each user, that maximizes the cascade in expectation. We study this problem under both non-adaptive and adaptive settings.
\begin{itemize}
\item In the non-adaptive setting, we have to commit the set of initial users and corresponding discounts all at once in advance. Although this setting is similar to traditional influence maximization problem, a unique challenge of our problem is to model the responses from the customer with respect to different discount rates. In particular, our model should be able to incorporate the following two constraints: (a) If a user receives multiple discounts, her decision on whether or not to accept the offer only depends on highest discount rate, and (b) Given that an initial user accepts a discount, she will accept all discounts with higher rate.   We prove that the utility function under this setting is monotone and submodular, which admits a $\frac{1}{2}(1 - 1/e)$-approximation algorithm.
\item Under the significantly more complex adaptive setting, the decision process is performed in a sequential manner, and each initial user that is picked provides the feedback on the discount rate, or, in other words, reveals whether or not she will adopt the discount. The action taken in each step depends on the actual cascade that happens in the previous steps. Therefore any feasible solutions are now policies instead of a fixed configuration. It was worth noting that our problem is closely related to adaptive/stochatic submodular maximization, however, there are two significant differences between the two: first of all, existing studies mainly assume that the cost of the action is known to the algorithm before the action is taken. However, this assumption is no longer valid under our setting, i.e, the actual amount of discount that has been delivered to a initial user depends on whether or not she accepts the offer; secondly, actions may incur non-uniform cost, this is obviously true since different discounts has different value. Unfortunately,  existing solutions can not handle the above two challenges. In this work, we prove that the utility function is adaptive monotone and adaptive submodular, this allows us to develop a novel greedy policy with bounded approximation ratio.
\end{itemize}

To the best of our knowledge, we are the first to systematically study the problem of discount allocation in both non-adaptive and adaptive settings. Under independent cascade model, which is one of the most commonly used models  in literature, we present a detailed analysis of the computational complexity of the problem. We propose a simple greedy algorithm with a constant approximation ratio in non-adaptive setting. We also develop a series of adaptive  policies with bounded approximation ratios under the adaptive setting.

\section{Related Work}
\label{sec:related}
 \cite{domingos2001mining} pointed out that data mining plays an important role in helping companies determine which potential customers to market to. Their work highlighted the importance of  a customer's network value which is derived from her influence on other users.
 We discuss related work on two related topics.  First topic is on the study of non-adaptive influence maximization problem:  given a social network, how to find a set of influential customers in order to trigger a large cascade of adoptions. Kempe et al. \citep{kempe2003maximizing} formulated the influence maximization problem under two diffusion models, namely Independent Cascade model and Linear Threshold model. Since then, considerable work \citep{chen2013information}\citep{leskovec2007cost}\citep{cohen2014sketch}\citep{chen2010scalable}\citep{chen2009efficient} has
been devoted to extending existing models to study influence maximization and its variants, but almost all these
works assume no uncertainty in the realization of seed set, i.e., any node that has been targeted will become seed or accept the offer immediately. Eftekhar et al. \citep{eftekhar2013information} relaxed this assumption by assuming that the probability that a user becomes a seed user is given and fixed.  In this work, we introduce the concept of adoption probability to capture the probability that a targeted user will accept the discount, based on the value of the discount and her own interest profile, i.e., the prior probability that the user will accept the discount in the absence of any social proof. Lastly, we mention the work by \cite{yangcontinuous} on continuous influence maximization. While they also study how to offer discounts to social users in order  to trigger large cascades, our settings are very different. First of all, they assume a continuous adoption function whereas we adopt a discrete function to capture the adoption probability. More importantly, from a theoretical perspective, we prove approximation guarantees for our approach. Secondly, their work mainly focuses on non-adaptive setting while our work covers both non-adaptive and adaptive settings. We propose a novel adaptive policy with bounded approximation ratio.

The second topic is on adaptive/stochastic submodular maximization \citep{golovin2011adaptive,badanidiyuru2016locally}. Our work departs from the body of work in this field in two ways: first of all, existing studies mainly assume that the cost of the action is known to the algorithm before the action is taken. However, this assumption is no longer valid under our setting, i.e, the actual amount of discount that has been delivered depends on whether or not the targeted user accepts the offer; secondly, actions may incur non-uniform cost, due to different discount values. Unfortunately,  existing solutions can not handle the above two challenges. In this work, we develop a novel greedy adaptive policy with constant approximation ratio.
\section{Network Model and Diffusion Process}
\label{sec:system}
A social network is modeled as a directed graph  $G = (V, E)$, where $V$ is a set of users
and $E$ is a set of social ties. Our model decomposes the cascade into two stages: \emph{seeding} and \emph{diffusion}. In the seeding stage, the initial set of users and corresponding discounts are decided for each initial user. The initial set of users who accept the discount act as starting points, called \emph{seed users},  in the diffusion stage. In the diffusion stage, starting from seed users,  the adoption propagates across the entire social network according to certain propagation model. Below we describe the details of these two stages.

\begin{itemize}
\item \textbf{Seeding stage:} Given the initial set of users and corresponding discount rate, each initial user decides whether or not to accept the offer. Although this decision may be affected by various factors, including the discount rate and user's attributes and behaviors (demographics, shopping history), it is reasonable to assume that the adoption probability of any initial user is monotonically increasing with respect to discount. Assume there are $m$ possible discount rates $\mathcal{D}=\{d_1, \cdots, d_m\}$, each user $v \in V$ is associated with an adoption probability function $p_v: d_i \rightarrow [0, 1]$, which models the probability that $v$ accepts the offer given a discount greater than or equal to $d_i$. It is clearly true that $p_v(d_i)\geq p_v(d_j)$ for any $v$ and discounts $d_i \geq d_j$. Estimating the adoption probability itself is a very important and challenging task, it has been well studied in the field of marketing. Interested readers may refer to \citep{brennan1995constructing}\citep{suh2004prediction} for further information. It is easy to verify that the above adoption model satisfies the following two conditions:
    \begin{definition} [Dominant Condition] If a user receives multiple discounts, her decision on whether or not to accept the offer only depends on the one with highest rate.
    \end{definition}
    \begin{definition}
   [Monotonic Condition] Given that a user has accepted a discount, she will accept any discount with higher rate.
  \end{definition}

     It was worth noting that all results derived in this work apply to any adoption model that is  \emph{dominant} and \emph{monotonic}.
\item \textbf{Diffusion stage:} Every initial user who decides to adopt the discount becomes the seed in the following diffusion stage, and starts to propagate the product to her neighbors across the social network. There are many ways to model the cascades in the social network, we adopt \emph{independent cascade model} \citep{kempe2003maximizing}, which is one of the most commonly used models, to model the diffusion dynamics in a social network. Under independent cascade model, each edge $(u, v)$ in the graph is associated with a propagation probability $p_{uv}$, which is the probability that node $u$ independently influences node $v$ in the next step after $u$ is influenced. Then given a set of seeds $U$, the independent cascade model works as follows: Let $U_t \subseteq V$ denote the set of influenced nodes at step $t$ with $U_0=U$. At step $t+1$, every user $v\in U_t$ may influence her out-neighbors $v\in V\setminus \cup_{0\leq i\leq t-1} U_i$ with an independent probability $p_{vu}$. This process ends at $t$ with $U_t=\emptyset$. The expected cascade of $U$, which is the expected number of influenced nodes given seed set $U$, is denoted as $I(U)$.
\end{itemize}

\section{Problem Formulation}
\subsection{Non-adaptive Setting}
Given the set of users $V$ and possible discount rates $\mathcal{D}$, define $\mathcal{H} \triangleq V\times \mathcal{D}$ as the solution space, probing \emph{seed-discount} (s-d) pair $h=\langle \mathbf{v}(h), \mathbf{d}(h) \rangle \in \mathcal{H} $ translates to offering $\mathbf{d}(h)\in \mathcal{D}$ to user $\mathbf{v}(h)\in V$. 
We use $\mathcal{S} \subseteq \mathcal{H}$ to denote a configuration of discounts assigned to a subset of initial users in $V$. Please note that it is feasible to assign multiple discounts to the same user, however, since her adoption decision only depends on the highest discount (due to \emph{dominant} condition), it suffice to use the highest discount as a representative. Let $d_\mathcal{S}[v]$ denote the highest discount assigned to $v$ under $\mathcal{S}$. Define $d_\mathcal{S}[w]=0$ if $w$ has not been selected as a initial user under $\mathcal{S}$, i.e, there is no $h\in \mathcal{S}$ with $\mathbf{v}(h)=w$. Given a social network $G=(V,E)$ and a configuration $\mathcal{S}$, the probability that a subset of users $U \subseteq V$ of the users is the seed set is
\[\Pr(U;V;\mathcal{S})=\prod_{u\in U} p_u(d_\mathcal{S}[u]) \prod_{v\in V\setminus U}(1- p_v (d_\mathcal{S}[v]))\]
As introduced earlier, we use $I(U)$ to denote the expected cascade under seed set $U$, then the expected cascade under configuration $\mathcal{S}$ is
 \[f(\mathcal{S})=\sum_{U\in 2^{V}}\Pr(U;V;\mathcal{S}) \cdot I(U)\]Now we define the non-adaptive discount allocation problem
(NDA) studied in this paper as follows. Given a social
network $G$, a propagation probability $p_{uv}$ for every edge $(u,v)$, a budget $B$, a seed probability function $p_u: d_i \rightarrow [0, 1]$ for every user $u$ and rate $d_i$, find the configuration $\mathcal{S}$ that is the optimal solution to the
following optimization problem.

\begin{center}
\framebox[0.78\textwidth][c]{
\enspace
\begin{minipage}[t]{0.55\textwidth}
\small
\textbf{NDA:}
\emph{Maximize $f(\mathcal{S})$}\\
\textbf{subject to:}
\begin{equation*}
\begin{cases}
\forall v\in \mathcal{V}, d_\mathcal{S}[v] \in \mathcal{D} \\
\sum_{v\in \mathcal{V}}  d_\mathcal{S}[v] \leq B

\end{cases}
\end{equation*}
\end{minipage}
}
\end{center}
\vspace{0.1in}
Although this setting is similar to traditional influence maximization problem, a unique challenge here is to model the responses from the user with respect to different discount rates. Notice that the budget constraint specified above ensures that the worst-case cost is bounded below by $B$. It is also reasonable to replace this hard constraint by other forms of constraint such as a soft constraint on the expected cost: $\sum_{v\in \mathcal{V}} d_\mathcal{S}[v]\cdot p_v (d_\mathcal{S}[v]) \leq B$. Our results are general enough to apply to both cases.

\begin{figure}[tbhp]
\begin{center}
\includegraphics[width=2.5in]{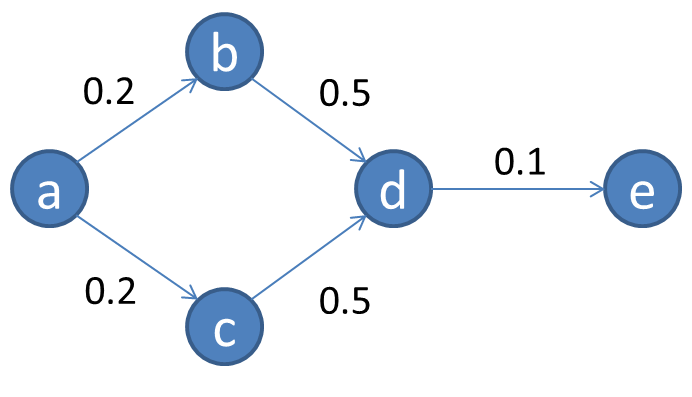}
\caption{A toy social network. Possible discounts: $\mathcal{D}=\{1, 2\}$; propagation probabilities are reported on edges; adoption probabilities: $\forall u\in \{a, b, c, d, e\}: p_u(1)=0.5, p_u(2)=1$; budget $B=2$.}
\label{fig:FSSILQI}
\end{center}
\end{figure}

For our example we use the toy social network in Fig. \ref{fig:FSSILQI}. In this example, there are five users $V=\{a, b, c, d, e\}$, adoption probabilities and propagation probabilities (on edges) are reported in the figure. Let us consider two ways of allocating discounts to users. The first one  is $\mathcal{S}_1=\{a, 2\}$, i.e.,  allocating $2$ to $a$: under $\mathcal{S}_1$, $a$ accepts the offer with probability $p_a(2)=1$; $b$ and $c$ will be influenced with probability $0.2$; $d$ will be influenced with probability $1-(1-0.2\times 0.5)^2=0.19$; $e$ will be influenced with probability $0.19\times 0.1=0.019$. The expected cascade size under $\mathcal{S}_1$ is $f(\mathcal{S}_1) = 1+0.2\times2 + 0.19 + 0.019=1.609$. Consider the second allocation $\mathcal{S}_2=\{a,1, b,1\}$, i.e., allocation $1$ to $a$ and $b$: under $\mathcal{S}_2$, $a$ (resp. $b$) accepts the offer with probability $p_a(1)=0.5$ (resp. $p_b(1)=0.5$); $c$ is influenced with probability $0.5\times 0.2 = 0.1$; $d$ is influenced with probability $1-(1-0.5)(1-0.1)=0.55$; $e$ is influenced with probability $0.55\times 0.1=0.055$. The expected cascade size under $\mathcal{S}_2$ is $f(\mathcal{S}_2) = 0.5+0.5+0.1+0.55+0.055=1.705$. Therefore $\mathcal{S}_2$ leads to larger expected cascade than $\mathcal{S}_1$.

\subsection{Adaptive Setting}
Different from the previous setting where we have to commit the set of initial users and corresponding discounts all at once in advance, under the adaptive setting, the decision process is performed in a sequential manner. Each initial user that is probed provides her feedback on the discount rate, or, in other words, reveals whether or not she will adopt the discount. The action taken in each step depends on the actual cascade that happens in the previous steps and remaining budget. Therefore any feasible solutions are now policies instead of a fixed configuration.
\begin{definition}[Seeding Realization]
For every configuration $h$,  $\mathbf{v}(h)$ is either in ``accept'' state ($h\rightarrow 1$) or in ``reject'' state ($h\rightarrow 0$), describing whether $\mathbf{v}(h)$ accepts $\mathbf{d}(h)$ or not. We represent the state of the seeding stage using function $\phi: \mathcal{H}\rightarrow [0,1]$, called \emph{seeding realization}.
\end{definition}
\begin{definition}[Diffusion Realization]
For every edge $(u,v)\in E$, it is either in ``live'' state or in ``blocked'' state. (describing whether the propagation through $(u,v)$ is a success or not). We represent the state of the diffusion stage using function $\psi: E \rightarrow [0,1]$, called \emph{diffusion realization}.
\end{definition}

After probing a s-d pair $h$ we are able to obtain the seeding realization, i.e., $\mathbf{v}(h)$ decides wether or not  to accept $\mathbf{d}(h)$. If $\mathbf{v}(h)$ decides to accept the discount, we further get to see the realization of the diffusion stage,  i.e., the status (live or dead) of all edges exiting $\mathbf{v}(h)$, for all nodes $v$ reachable from $\mathbf{v}(h)$ via live edges social network i.e., reachable from $\mathbf{v}(h)$ under $\psi$. After each attempt of probing, our observations so far can be represented as a partial realization $\langle \phi_p, \psi_p\rangle$.

\begin{definition}[Adaptive Policy]
 We define our adaptive policy $\pi: \langle\phi_p, \psi_p\rangle\rightarrow \mathcal{H}$, which is a function from the current ``observation'' $\langle\phi_p, \psi_p\rangle$ to $\mathcal{H}$, specifying which s-d pair to pick next under a particular set of observations, e.g., $\pi$ chooses s-d pair given s-d pairs have been probed so far, and the resulting cascade.
 \end{definition}

 Assume there is a known prior probability distribution
$p (\phi) := P [\Phi = \phi]$ (and $p (\psi) := P [\Psi = \psi]$ resp.) over seeding  realizations (and diffusion realizations resp.).  Given a realization $\langle\phi, \psi\rangle$, let $\mathcal{H}(\pi; \phi, \psi)$ denote all s-d pairs picked by $\pi$ under $\langle\phi, \psi\rangle$, and $c(\mathcal{H}(\pi; \phi, \psi))$ denote the total amount of discount that have been delivered by $\pi$ under $\langle\phi, \psi\rangle$. The expected cascade of a policy $\pi$ is $f(\pi) = \mathbb{E}[f(\mathcal{H}(\pi; \Phi, \Psi)| \Phi, \Psi)]$ where the
expectation is taken with respect to $p(\phi)$ and $p(\psi)$. The goal of the Adaptive Coupon Distribution (ACD) problem is to find
a policy $\pi$ such that
\begin{center}
\framebox[0.78\textwidth][c]{
\enspace
\begin{minipage}[t]{0.55\textwidth}
\small
\textbf{ACD:}
\emph{Maximize $f(\pi)$}\\
\textbf{subject to:}
\begin{equation*}
c(\mathcal{H}(\pi; \phi, \psi)) \leq B, \forall \phi, \psi
\end{equation*}
\end{minipage}
}
\end{center}
\vspace{0.1in}

\begin{figure}[tbhp]
\begin{center}
\includegraphics[width=6in]{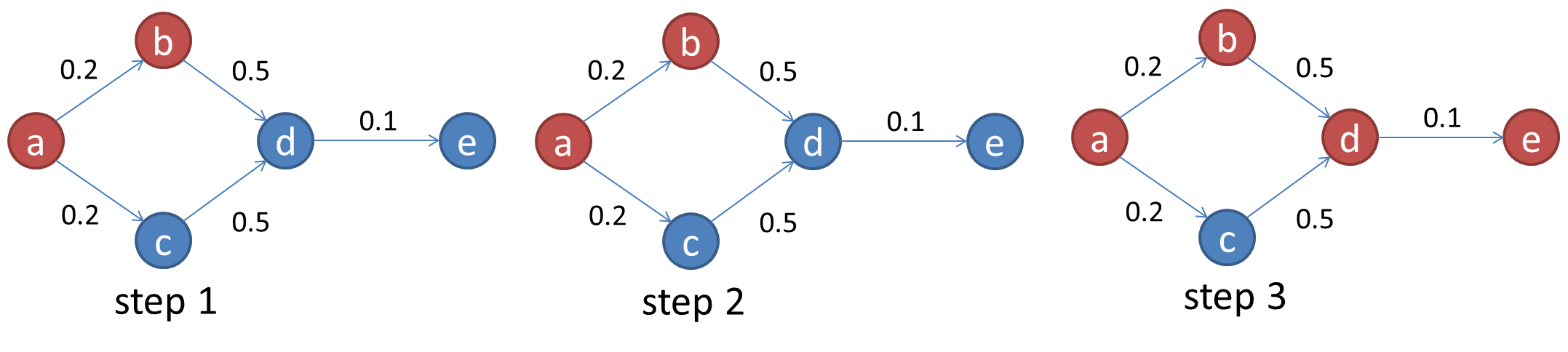}
\caption{Step 1: pick $\langle a, 1 \rangle$; step2: pick $\langle c, 1 \rangle$; step 3: pick $\langle d, 1 \rangle$.}
\label{fig:2}
\end{center}
\end{figure}

For our example we use the toy social network in Fig. \ref{fig:2}, all settings of this example are identical to Fig. \ref{fig:FSSILQI}. Consider a possible adaptive policy as follows: We first probe $\langle a, 1 \rangle$, that is offering discount $1$ to user $a$, and observe the following partial seeding realization: $\phi(\langle a, 1 \rangle)=1$, i.e., $a$ accepts the offer; and partial diffusion realization: $\psi(ab)=1$ and $\psi(ac)=0$, i.e., user $b$ is influenced by $a$, user $c$ has not been influenced. In the second step where the remaining budget is 1, we probe $\langle c, 1 \rangle$, that is offering discount $1$ to user $c$ and obtain the following seeding realization: $\phi(\langle c, 1 \rangle)=0$, i.e., user $c$ does not accept the offer. In the third step where the remaining budget is still 1, we probe $\langle d, 1 \rangle$, that is offering discount $1$ to user $d$, and observe that $\phi(\langle d, 1 \rangle)=1$, i.e., user $d$ has accepted this offer, and $\psi(de)=1$, i.e., $e$ has been influenced by $d$. Thus the number of influenced users is 4.

\section{Non-Adaptive Discount Allocation Problem}
In this section, we study non-adaptive discount allocation problem. Our main idea is to prove that the influence function $f(\cdot)$ is monotone and submodular, exploiting these properties, a simple greedy algorithm can achieve $\frac{1}{2}(1-1/e)$ approximation ratio.

\begin{definition}[Submodular function] Consider an arbitrary function $z(\cdot)$ that maps subsets of a finite ground set $\Omega$ to non-negative real numbers. We say that $z(\cdot)$ is submodular if it satisfies a natural ``diminishing returns" property: the marginal gain from adding an element to a set $S$ is at least as high as the marginal gain from adding the same element to a superset of $S$. Formally, a submodular function satisfies the follow property: For every $X, Y \subseteq \Omega$ with $X \subseteq Y$ and every $x \in \Omega \backslash Y$, we have that \[z(X\cup \{x\})-z(X)\geq z(Y\cup \{x\})-z(Y)\]
We say a submodular function $z$ is monotone if $z(X) \leq z(Y)$ whenever $X \subseteq Y$.
\end{definition}

To prove that the influence function $f(\cdot)$ is monotone and submodular, we need to focus on the value of $f(\mathcal{S} \cup h) - f(\mathcal{S})$, for arbitrary sets $\mathcal{S}\subseteq \mathcal{H}$ and s-d pair $h$. Since the increase of the above expression is very difficult to analyze directly, we take an equivalent view of the cascade process by decomposing it into two realizations:

\begin{enumerate}
\item \underline{seeding realization  $\phi$.} Consider a point in the seeding stage when $h=\langle v, d \rangle$ is probed, i.e,  node $v$ has been offered discount $d$, $v$ accepts  offer and becomes the seed with probability $p_v(d)$. We can view the outcome of this random event as being determined by comparing $p_v(d)$ with a random number $g$ that is uniformly selected from $[0, 1]$: if $p_v(d)\geq g$, then $h=\langle v, d \rangle$ is declared to be \emph{valid} under $\phi$.
\item \underline{diffusion realization  $\psi$.} We follow the same approach introduced in \citep{kempe2003maximizing} to model the diffusion dynamics. In particular, for each edge $(u, v)$ in the graph, a coin of
bias $p_{uv}$ is flipped at the very beginning of the process. The edges for which the coin flip indicated an
 successful activation are declared to be \emph{live} in $\psi$; the remaining edges are declared to be \emph{blocked} in $\psi$. As described in detail later, a user is influenced if it can be reached from some seed through a path consisting of live edges.
\end{enumerate}

Similar to the diffusion realization, where the coin can be flipped at the beginning of the process, the seeding realization can also be performed ahead of the actual seeding stage. In particular, for each node $u$, a random number $g_u$ is uniformly selected from $[0, 1]$: if there exists a $h\in \mathcal{H}$ with $\mathbf{v}(h)=u$ and $\mathbf{d}(h)\geq g_u$, then $\mathbf{v}(h)$ is declared to be \emph{live} under discount $\mathbf{d}(h)$. Given a configuration $\mathcal{S} \subseteq \mathcal{H}$ and a seeding realization $\phi$,  we say a node $\mathbf{v}(h)$ is \emph{live} under $\mathcal{S}$ and $\phi$ if and only if there exists $h\in \mathcal{S}$ such that $\mathbf{v}(h)$ is live with discount $\mathbf{d}(h)$. With both seeding and diffusion realizations are performed in advance,
it is easy to determine the subset of nodes that can be influenced at the end of the cascade process:

\begin{claim} Given a configuration $\mathcal{S}$ and realization $\langle\phi, \psi\rangle$, a node $u$ ends up influenced if and only if $u$ is \underline{live} or there is a path from some \underline{live} node to $u$ consisting entirely of \underline{live} edges.
\end{claim}

\begin{theorem}
The influence function $f(\cdot)$ is submodular.
\end{theorem}
\emph{Proof:} We first prove that for each fixed realization $\langle\phi,\psi\rangle$, the function $f(\cdot)$ is submodular. We use  $f(\cdot|{\phi,\psi })$ to denote the (deterministic) influence function under a fixed realization $\langle\phi,\psi\rangle$. Let $\mathcal{A}$ and $\mathcal{B}$ be two configurations such that $\mathcal{A} \subseteq \mathcal{B}$, and consider the value $f(\mathcal{A}\cup\{h\}|{\phi,\psi})-f(\mathcal{A}|{\phi,\psi})$, this is the number of users that can be reached by live users in $\mathcal{A}\cup\{h\}$ but not in $\mathcal{A}$. This number is at least as large as the number of users that can be reached by live users in $\mathcal{B}\cup\{h\}$ but not in $\mathcal{B}$. It implies that $f(\mathcal{A}\cup\{h\}|{\phi,\psi })-f(\mathcal{A}|{\phi,\psi}) \geq f(\mathcal{B}\cup\{h\}|{\phi,\psi })-f(\mathcal{B}|{\phi,\psi })$
Then due to the fact that  a non-negative linear combination of submodular functions is also submodular, $f(\cdot)$ is also submodular. $\Box$

We next propose a hill-climbing algorithm (Algorithm \ref{alg:weight}) with a constant approximation ratio. Our algorithm computes two candidate sets: The first candidate set contains a single s-d pair which can
maximize the expected cascade: $\mathcal{S}_{1} = \{h^*\}$ where $h^*=\arg\max_{h}f(\{h\})$; the second candidate
solution $\mathcal{S}_2$ is computed by greedy algorithm in which
we always add $h$ that can maximize the expected
incremental marginal gain with respect to the discount rate: $\frac{f(\mathcal{S}_2\cup \{h\})}{\mathbf{d}(h)}$ until
the budget constraint is violated. Then we choose the better
one as the final output $\mathcal{S}$. It was worth noting that computing $f(\mathcal{S})$ is \#P-hard \citep{chen2010scalable}, and is typically approximated by numerous Monte Carlo simulations, however, running such simulations are extremely time consuming. Instead, we can leverage a FPRAS (Fully Polynomial Randomized Approximation Scheme) \citep{long2011minimizing} for estimating $f(\mathcal{S})$. Based on \citep{khuller1999budgeted} and the submodularity of $f(\cdot)$, we can prove that our greedy algorithm
achieves $\frac{1}{2}(1-e^{-1})$-approximation.

\begin{algorithm}
\caption{Hill-Climbing Algorithm}
\begin{algorithmic}[1]
\STATE $\mathcal{S}_{1}:h^*=\arg\max_{h}f(\{h\})$;  $\mathcal{S}_{2}:=\emptyset$;
\WHILE {$\mathbf{d}(\mathcal{S}_2)\leq B$}
\STATE $\mathcal{S}_2 \leftarrow \arg\max_{h \in \mathcal{H}} \frac{f(\mathcal{S}_2\cup \{h\})}{\mathbf{d}(h)}$; $\mathcal{H}=\mathcal{H}\setminus \{h\}$
\ENDWHILE
\STATE return $\arg\max_{\mathcal{S}\in \{\mathcal{S}_{1}, \mathcal{S}_{2}\}}f(\mathcal{S})$
\end{algorithmic}
\label{alg:weight}
\end{algorithm}

\begin{theorem}
Algorithm \ref{alg:weight} achieves approximation ratio $\frac{1}{2}(1-e^{-1})$.
\end{theorem}

\section{Adaptive Discount Allocation Problem}
We next study the adaptive discount allocation problem. In the rest of this paper, let $f(\pi|\phi)$ (resp. $f(\pi|\psi)$), as a shorthand notation for $f(\pi|\phi, \Psi)$ (resp. $f(\pi|\Phi, \psi)$), denote the expected cascade of $\pi$ under a fixed seeding realization $\phi$ (resp. a fixed diffusion realization $\psi$).
\subsection{Greedy Policy}
We first propose a adaptive greedy policy with bounded approximation ratio.
\subsubsection{Policy Description}
 Our greedy policy is performed in a sequential manner as follows: In each round, we probe the s-d pair, say $h^*$, that maximizes ratio of conditional expected marginal benefit to cost. If $\mathbf{v}(h^*)$ accepts $\mathbf{d}(h^*)$, we deduct the discount from the budget and remove $\mathbf{v}(h^*)$ from consideration in the following rounds. Otherwise, if $\mathbf{v}(h^*)$ turns down the discount, we simply discard $h^*$ and move to the next round. This process iterates until either the budget is used up or all nodes have been probed at the highest discount rate. The detailed description is listed in Algorithm \ref{alg:LPP1}: Given partial diffusion realization $\psi_p$, let $\mathrm{dom}(\psi_p)$ denote all influenced users under $\psi_p$,  we use $G [V \setminus \mathrm{dom}(\psi_p)]$ to denote the induced graph of $V\setminus  \mathrm{dom}(\psi_p)$. Let $\Delta(h|\psi_p)=I_{G [V \setminus \mathrm{dom}(\psi_p)]}(\mathbf{v}(h))$ denote the expected marginal benefit of $\mathbf{v}(h)$ in $G [V \setminus \mathrm{dom}(\psi_p)]$ \underline{conditioned on $\mathbf{v}(h)$ has become the seed}.
At beginning of each round, we check whether or not there exists $h \in \mathcal{H}$ such that $\mathbf{d}(h)\leq B$. If so,  probe $h^*=\arg\max_{h \in \mathcal{H}}\Delta(h|\psi_p)/\mathbf{d}(h)$ subject to $\mathbf{d}(h)\leq B$. Depending on the response from the $\mathbf{v}(h^*)$, we either commit $\mathbf{v}(h^*)$ and deduct $\mathbf{d}(h^*)$ from the remaining budget or skip to the next round.
\begin{algorithm}[hptb]
\caption{Greedy Policy}
\label{alg:LPP1}
\begin{algorithmic}[1]
\STATE $\mathcal{S}=\emptyset$
\WHILE {$B \geq 0$}
\IF {there exists $h \in \mathcal{H}$ such that $\mathbf{d}(h)\leq B$}
\STATE probe $h^*=\arg\max_{h \in \mathcal{H}}\Delta(h|\psi_p)/\mathbf{d}(h)$ subject to $\mathbf{d}(h)\leq B$
\IF {$\mathbf{v}(h^*)$ accepts $\mathbf{d}(h^*)$}
\STATE $\mathcal{S}\leftarrow \mathcal{S}\cup \{h^*\}$; $B \leftarrow B-\mathbf{d}(h^*)$;
\STATE $\mathcal{H}\leftarrow \mathcal{H}\setminus \{h|\mathbf{v}(h)=\mathbf{v}(h^*)\}$;
\STATE update $\psi_p$;
\ELSE
\STATE $\mathcal{H}\leftarrow \mathcal{H}\setminus \{h^*\}$
\ENDIF
\ELSE
\STATE break.
\ENDIF
\ENDWHILE
\RETURN $\mathcal{S}$
\end{algorithmic}
\end{algorithm}

We next walk through this greedy policy using Fig. \ref{fig:2}. We first probe $\langle a, 1 \rangle$ since it has the highest benefit-to-cost ratio $\frac{I(\{a\})}{1}$. We observe that $a$ accepts the offer and successfully influenced $b$. Conditioned on the above observation, we next probe $\langle c, 1 \rangle$ because it has the highest benefit-to-cost ratio in $G[V \setminus\{a, b\}]$, and we observe that $c$ turns down the offer. Then we probe $\langle d, 1 \rangle$ which has the highest benefit-to-cost ratio in $G[V\setminus\{a, b, c\}]$, and observe that $d$ has accepted the offer and successfully influenced $e$.
\subsubsection{Performance Analysis}
We first study a relaxed version of ADA by assuming that the seeding realization is pre-known. Given a seeding realization $\phi$, each node $v$ is associated with a minimum discount $d_v$ at which $v$ can become the seed. If $d_v$ will never become the seed under $\phi$, then $d_v=\infty$. Since it is meaningless to probe a user $v$ with any discount lower or higher than $d_v$, we use $\mathcal{H}^{\mathrm{relaxed}}=\{\langle v, d_v\rangle|v\in V\}$ to denote the refined solution space under this relaxed setting. Consider a greedy policy $\pi^{\mathrm{greedy}}_{\mathrm{relaxed}}$ as follows: In each round,  update the partial diffusion realization $\psi_p$, and probe $h\in \mathcal{H}^{\mathrm{relaxed}}$ that maximizes $\Delta(h|\psi_p)/d_{\mathbf{v}(h)}$. This process iterates until either the budget is used up or all users have been probed.
\begin{lemma}
\label{lem:main}
 Under the relaxed setting, given any seeding realization $\phi$, the greedy policy $\pi^{\mathrm{greedy}}_{\mathrm{relaxed}}$ obtains at least $(1-e^{-(B-d_{\max})/B})$ of the value of the best policy $\pi^*_{\mathrm{relaxed}}$.
\[f(\pi^{\mathrm{greedy}}_{\mathrm{relaxed}}|\phi) \geq (1-e^{-(B-d_{\max})/B})f(\pi^*_{\mathrm{relaxed}}|\phi)\]
\end{lemma}
\emph{Proof:} Consider two types of seeding realizations depending on the output of $\pi^{\mathrm{greedy}}$: 1. $\phi^a$: all users except those whose minimum discount is $\infty$ have become the seeds; 2.  $\phi^b$: there exists some user with finite minimum discount that can not be included in the solution due to limited budget.

The first case is trivial, because both $\pi^{\mathrm{greedy}}_{\mathrm{relaxed}}$ and $\pi^*_{\mathrm{relaxed}}$ must select all possible seeds,
\begin{equation}
\label{eq:a}
f(\pi^{\mathrm{greedy}}_{\mathrm{relaxed}}|\phi^a) = f(\pi^*_{\mathrm{relaxed}}|\phi^a)
\end{equation}

We next focus on the second case:
In Theorem 26 \citep{golovin2011adaptive}, it has been proved that\[f(\pi_{[l]}) \geq (1-e^{-l/k})f(\pi^*_{[k]})\]
where $\pi_{[l]}$ denotes the greedy policy with \emph{expected} budget $l$ and $\pi^*_{[k]}$ denotes the best policy with \emph{expected} budget $k$. The above result is applicable to the traditional influence maximization problem without considering seeding stage. Taking seeding stage into account, it is easy to extend this result to show that $f(\pi^{\mathrm{greedy}}_{\mathrm{relaxed}}|\phi^b) \geq f(\pi_{[B-d_{\max}]}|\phi^b)$ and $f(\pi^*_{\mathrm{relaxed}}|\phi^b) \geq f(\pi_{[B]}|\phi^b)$, this is because the actual cost of $\pi^{\mathrm{greedy}}_{\mathrm{relaxed}}$ is lower bounded by $B-d_{\max}$, and the actual cost of $\pi^*$ is upper bounded by $B$. It follows that
\begin{equation}
\label{eq:b}
f(\pi^{\mathrm{greedy}}_{\mathrm{relaxed}}|\phi^b) \geq (1-e^{-(B-d_{\max})/B})f(\pi^*_{\mathrm{relaxed}}|\phi^b)
\end{equation}
Eqs. (\ref{eq:a}) and (\ref{eq:b}) together imply that, for all seeding realization $\phi$, we have
\[f(\pi^{\mathrm{greedy}}_{\mathrm{relaxed}}|\phi) \geq (1-e^{-(B-d_{\max})/B})f(\pi^*_{\mathrm{relaxed}}|\phi)\]
$\Box$

Surprisingly, we next show that given any realization $\langle \phi,\psi \rangle$, the outputs of $\pi^{\mathrm{greedy}}$ and $\pi^{\mathrm{greedy}}_{\mathrm{relaxed}}$ are identical.
\begin{lemma}
\label{lem:1}
Given any realization $\langle \phi,\psi \rangle$, the outputs of $\pi^{\mathrm{greedy}}$ and $\pi^{\mathrm{greedy}}_{\mathrm{relaxed}}$ are identical.
\end{lemma}
\emph{Proof:} Recall that under $\pi^{\mathrm{greedy}}$, we probe  $h^*=\arg\max_{h \in \mathcal{H}}\Delta(h|\psi)/\mathbf{d}(h)$ in each round. Depending on the response from the $\mathbf{v}(h^*)$, we either commit $h^*$ or skip to the next round. We next show that by following $\pi^{\mathrm{greedy}}$, 
if $\mathbf{v}(h^*)$ accepts $h^*$, then $\mathbf{d}(h^*)=d_{\mathbf{v}(h^*)}$, i.e., $\pi^{\mathrm{greedy}}$ will never probe a user with discount higher than her minimum discount. First, it is trivial to verify $d_{\mathbf{v}(h^*)}\leq \mathbf{d}(h^*)$ due to the definition of $d_{\mathbf{v}(h^*)}$. We can also prove that otherwise, we can probe $\mathbf{v}(h^*)$ with discount $d_{\mathbf{v}(h^*)}$ which is lower than $\mathbf{d}(h^*)$, leading to higher benefit-to-cost ratio. This contradicts to the assumption that $h^*=\arg\max_{h \in \mathcal{H}}\Delta(h|\psi)/\mathbf{d}(h)$.

We are now ready to prove that the outputs of $\pi^{\mathrm{greedy}}$ and $\pi^{\mathrm{greedy}}_{\mathrm{relaxed}}$ are identical. Let $S_t$ denote the first $t$ seeds committed by $\pi^{\mathrm{greedy}}$, we prove this result through induction on $t$. The basic case when $t=0$, i.e., $S_0=\emptyset$ is trivial. As proved above, when $\mathbf{v}(h^*)$ accepts $h^*$,  we have $d_{\mathbf{v}(h^*)}=\mathbf{d}(h^*)$. Therefore, the first seed that $\pi^{\mathrm{greedy}}$ and $\pi^{\mathrm{greedy}}_{\mathrm{relaxed}}$ commit must be some user $v$ that maximizes $I(\{v\})/d_v$. Assume by contradiction that $\pi^{\mathrm{greedy}}$ commits some user other than $v$, then probing $v$ with discount less than or equal to $d_v$ leads to a higher  benefit-to-cost ratio. This causes contradiction to the design of $\pi^{\mathrm{greedy}}$. Assume the first $k$ users committed by $\pi^{\mathrm{greedy}}$ and $\pi^{\mathrm{greedy}}_{\mathrm{relaxed}}$ are identical, we next prove that the $(k+1)$-th seed is also identical. Given the partial observation $\psi_p$ after the first $k$ seeds have been committed, the next s-d pair that $\pi^{\mathrm{greedy}}$ commits must be some $h$ that maximizes $\Delta(h|\psi_p)/d_{\mathbf{v}(h)}$, otherwise, probing $\mathbf{v}(h)$ with discount less than or equal to $d_{\mathbf{v}(h)}$ leads to a higher  benefit-to-cost ratio. Therefore, $\pi^{\mathrm{greedy}}$ and $\pi^{\mathrm{greedy}}_{\mathrm{relaxed}}$ commit the same set of seeds. This finishes the proof of this lemma. $\Box$

\begin{theorem}
\label{thm:main}
The greedy policy $\pi^{\mathrm{greedy}}$ obtains at least $(1-e^{-(B-d_{\max})/B})$ of the value of the best policy $\pi^*$.
\[f(\pi^{\mathrm{greedy}}) \geq (1-e^{-(B-d_{\max})/B})f(\pi^*)\]
\end{theorem}
\emph{Proof:}  We have proved that $f(\pi^{\mathrm{greedy}}_{\mathrm{relaxed}}|\phi) \geq (1-e^{-(B-d_{\max})/B})f(\pi^*_{\mathrm{relaxed}}|\phi)$ for all $\phi$. Based on the definition of $f(\pi^*_{\mathrm{relaxed}}|\phi)$, we have $f(\pi^*_{\mathrm{relaxed}}|\phi)\geq f(\pi^*|\phi)$.
Thus, \[f(\pi^{\mathrm{greedy}}_{\mathrm{relaxed}}|\phi) \geq (1-e^{-(B-d_{\max})/B})f(\pi^*_{\mathrm{relaxed}}|\phi)\geq (1-e^{-(B-d_{\max})/B})f(\pi^*|\phi)\]
Then together with Lemma \ref{lem:1}, we have
\[f(\pi^{\mathrm{greedy}}|\phi) = f(\pi^{\mathrm{greedy}}_{\mathrm{relaxed}}|\phi) \geq (1-e^{-(B-d_{\max})/B})f(\pi^*_{\mathrm{relaxed}}|\phi)\geq (1-e^{-(B-d_{\max})/B})f(\pi^*|\phi)\]
It follows that
\[f(\pi^{\mathrm{greedy}}|\Phi) \geq (1-e^{-(B-d_{\max})/B})f(\pi^*|\Phi)\]
This finishes the proof of this theorem. $\Box$

We observe that the above greedy heuristic has an unbounded approximation ratio in the worst case.  Consider, for example, a network consists of $n$ nodes: one isolated node $x$ and  a clique with size $n-1$. There are two discount rates $\mathcal{D}=\{1/n, 1\}$, and budget is $1$. We further set $p_x(1/n) = p_x(1) = 1$, and $p_y(1/n)=0, p_y(1)=1, \forall y \in V\setminus \{x\}$. Assume $I(\{x\})=1$ and $I(\{y\})=n-1, \forall y \in V\setminus \{x\}$. By following the heuristics, we will provide discount $1/n$ to $x$, reaching $1$ user in expectation. However, the best strategy is to provide discount $1$ to any user except $x$,  reaching cascade $n-1$ in expectation. Thus the approximation ratio in this case is as large as $n$.

\subsection{Enhanced Greedy Policy}
We next show that a small modification to the heuristic can achieve an approximation ratio independent of $n$.
Our policy (Algorithm \ref{alg:LPP2}) first computes two candidate solutions: The first candidate solution contains a single node $v^*$ which can maximize the expected cascade given that $v^*$ has become the seed: $v^* = \arg\max_{v\in V}I(\{v\})$; the second candidate solution is computed by the greedy policy.  Then we choose the  one that leads to larger expected cascade as the final solution.

\begin{algorithm}[hptb]
\caption{Enhanced Greedy Policy $\pi^{\mathrm{enhanced}}$}
\label{alg:LPP2}
\begin{algorithmic}[1]
\STATE Let $v^* = \arg\max_{v\in V}I(\{v\})$;
\IF {$p_{v^*}(d_{\max})\cdot I(\{v^*\})>f(\pi^{\mathrm{greedy}})$}
\STATE offer $d_{\max}$ to $v^*$;
\ELSE
\STATE call $\pi^{\mathrm{greedy}}$
\ENDIF
\end{algorithmic}
\end{algorithm}

\begin{theorem}
Let $v^* = \arg\max_{v\in V}I(\{v\})$ and $d_{\max}$ denote the highest discount rate, our enhanced greedy policy $\pi^{\mathrm{enhanced}}$ obtains at least $p_{v^*}(d_{\max})\cdot (1-e^{-1})/2$ of the value of the best policy $\pi^*$ for the Adaptive
Coupon Distribution problem with the independent cascade model.
\[f(\pi^{\mathrm{enhanced}}) \geq p_{v^*}(d_{\max})\cdot \frac{(1-e^{-1})}{2}\cdot f(\pi^*)\]
\end{theorem}
\emph{Proof:} Similar to proof of Theorem \ref{thm:main}, we consider two types of realizations $\phi^a$ and  $\phi^b$.  

We first prove that the following equality holds under $\phi^a$: $f(\pi^{\mathrm{greedy}}|\phi^a)=f(\pi^*|\phi^a)$. This is because $\pi^{\mathrm{greedy}}$ and $\pi^*$ must select the same set of seeds due to the definition of $\phi^a$. It follows that
\begin{equation}
\label{eq:1}
f(\pi^{\mathrm{greedy}}|\phi^a)+ I(\{v^*\}) \geq f(\pi^*|\phi^a)
\end{equation}

Next, consider $\phi^b$, we have
\begin{equation}
\label{eq:2}
f(\pi^{\mathrm{greedy}}|\phi^b) + I(\{v^*\}) \geq f(\pi_{[B]}|\phi^b) \geq (1-e^{-1})f(\pi^*|\phi^b)
\end{equation}
The first inequality is due to the adaptive submodularity of $I(\cdot)$,  and the second inequality is based on Theorem \ref{thm:main}.
Eqs. (1) and (2) together imply that, under any realization $\phi$:
\[f(\pi^{\mathrm{greedy}}|\phi) + I(\{v^*\})\geq (1-e^{-1})f(\pi^*|\phi)\]
It follows that
\[f(\pi^{\mathrm{greedy}}) + I(\{v^*\}) \geq (1-e^{-1})f(\pi^*)\]
We further have
\[f(\pi^{\mathrm{greedy}}) + p_v(d_{\max})\cdot I(\{v^*\}) \geq p_v(d_{\max}) \cdot (1-e^{-1})f(\pi^*)\]
Then
\[\max\{f(\pi^{\mathrm{greedy}}), p_v(d_{\max})\cdot I(\{v^*\})\} \geq p_v(d_{\max})\cdot \frac{(1-e^{-1})}{2}\cdot f(\pi^*)\]

%
%
%
$\Box$

\begin{corollary}
Assume $p_{v^*}(d_{\max})=1$, the enhanced policy $\pi^{\mathrm{enhanced}}$ obtains at least $(1-e^{-1})/2$ of the value of the best policy $\pi^*$ for the Adaptive Discount Allocation problem with the independent cascade model.
\[f(\pi^{\mathrm{enhanced}}) \geq \frac{(1-e^{-1})}{2}\cdot f(\pi^*)\]
\end{corollary}

To further enhance the performance of Algorithm \ref{alg:LPP2}, we develop another heuristic as listed Algorithm \ref{alg:LPP5}. Algorithm \ref{alg:LPP5} is a natural extension of Algorithm \ref{alg:LPP2}: In case $v^*$ does not accept $d_{\max}$, we remove $v^*$ from consideration of initial nodes and apply Algorithm \ref{alg:LPP2} to the remaining nodes. Otherwise, if $v^*$ accepts $d_{\max}$, we update the budget and diffusion realization, and apply Algorithm \ref{alg:LPP2} to the remaining graph $G \setminus \mathrm{dom(\psi_{t})}$ subject to updated budget. Clearly, the cascade gained from this heuristic is no smaller than the one gained from  Algorithm \ref{alg:LPP2}, thus we have
\begin{theorem}
Algorithm \ref{alg:LPP5} obtains at least $p_{v^*}(d_{\max})\cdot (1-e^{-1})/2$ of the value of the best policy $\pi^*$ for the Adaptive
Discount Allocation problem with the independent cascade model.
\end{theorem}

\
\begin{algorithm}[hptb]
\caption{A Heuristic}
\label{alg:LPP5}
\begin{algorithmic}[1]
\WHILE {$V\neq \emptyset$}
\STATE Let $v^* = \arg\max_{v\in V}I_{G \setminus \mathrm{dom}(\psi_t)}(v)$;
\IF {$p_{v^*}(d_{\max})\cdot I(\{v^*\})>f(\pi^{\mathrm{greedy}}|\psi_t)$}
\STATE offer $d_{\max}$ to $v^*$
\IF {$v^*$ accepts $d_{\max}$}
\STATE $\mathcal{S}\leftarrow \mathcal{S} \cup \langle v^*, d_{\max}\rangle$; update the diffusion realization to $\psi_{t+1}$;
\ELSE
\STATE $V\leftarrow V\setminus \{v^*\}; \psi_{t+1}\leftarrow \psi_{t}$
\ENDIF
\ELSE
\STATE call $\pi^{\mathrm{greedy}}$ on $G \setminus \mathrm{dom}(\psi_t)$; break
\ENDIF
\ENDWHILE
\RETURN $\mathcal{S}$
\end{algorithmic}
\end{algorithm}

\bibliographystyle{ijocv081}
\bibliography{reference}




\end{document}